\documentclass[aps,prl,preprint,groupedaddress]{revtex4}
\usepackage{graphicx}

\begin{document}


\title{Faraday waves under perpendicular electric field}


\author{Ra\'ul Fern\'andez-Mateo}
\author{Alberto T. P\'erez}
\email[]{alberto@us.es}
\affiliation{Departamento de Electr\'onica y Electromagnetismo. Universidad de Sevilla}


\date{\today}

\begin{abstract}

The dispersion relation of vertically oscillating fluid surfaces has been a subject extensively studied in the past, as well as surface instabilities produced by electrohydrodynamic (EHD) waves in similar configurations. In the present work it is studied the unification of both effects and its consequences to the instability of the surface. Given the versatility of the electromagnetic fields, a possible application to the phenomenon of walking droplets is suggested.



\end{abstract}


\maketitle

\section{Introduction}

Since the pioneering works by Melcher and Taylor \cite{Melcher1961,Melcher1963,Taylor1965}, electrohydrodynamic (EHD) surface waves and instabilities have been the subject of many studies. Taylor and McEwan \cite{Taylor1965} studied the stability of a perfect conductor liquid layer under a direct current (DC) electric field. Melcher  \cite{Melcher1961,Melcher1963} provided the dispersion relation of EHD waves under  DC fields for several configurations, including two dielectric liquid layers with or without surface charge density at the interface. The effect of a finite conductivity was studied by Melcher and Smith \cite{Melcher1969}. Further extensions, including viscosity, an other physical effects have been studied since then \cite{Kath1977,Chu1989,Castellanos1992}. Non-linear waves were considered by Castellanos and Gonz\'alez \cite{Castellanos1998}. 

The use of AC electric fields of low frequency opens the possibility of electric parametric instabilities. These were studied by Briskman and Shaidurov \cite{Briskman1968} and Yih \cite{Yih1968}. They obtained a Mathieu equation, similar to the one appearing in the study of Faraday waves. A similar behavior may be found in cylindrical geometry in the study of liquid jets under AC electric fields \cite{Gonzalez1997}. A recent work is that of Ward et al \cite{Ward2019}.

On the other hand, surface instabilities produced in an air-liquid interface (first observed by Faraday in 1831, and named after him) were first studied theoretically by Benjamin and Ursell \cite{Benjamin1954}, and its resulting dispersion relation is governed by a \textsl{Mathieu equation}, whose stability analysis (see \cite{Mclachlan1947} for a complete mathematical description) yields the relation between parameters (mainly frequency and amplitude of oscillations) for the surface to become unstable. Later works which take into account the presence of viscosity \citep{Kumar1996} arrive to similar conclusions in a more precise description that move away from Mathieu's. 


In the present work we introduce the unification of the dispersion relation due to vertical forcing and the one given by the application of a vertical electric field. It will be shown that the instability can come from any of the two phenomena. 

The motivation of this work is to discuss the possible applications this unification might provide on the phenomenon of walking droplets. Until now, droplet guidance and confinement is carried out by modifying the bottom topology of the fluid container (see for example \cite{Protiere2006,Eddi2009}). It is here suggested that with the convenient application of an electric field, this guidance could also be achieved, opening a new way of controlling the droplets. This would be indeed a great advantage since it would be no longer needed to modify the depth of the container and with that the configuration of the whole system, making possible the existence of more versatile setups.

\section{Position of the problem}

\begin{figure}
\includegraphics[width=7.5cm]{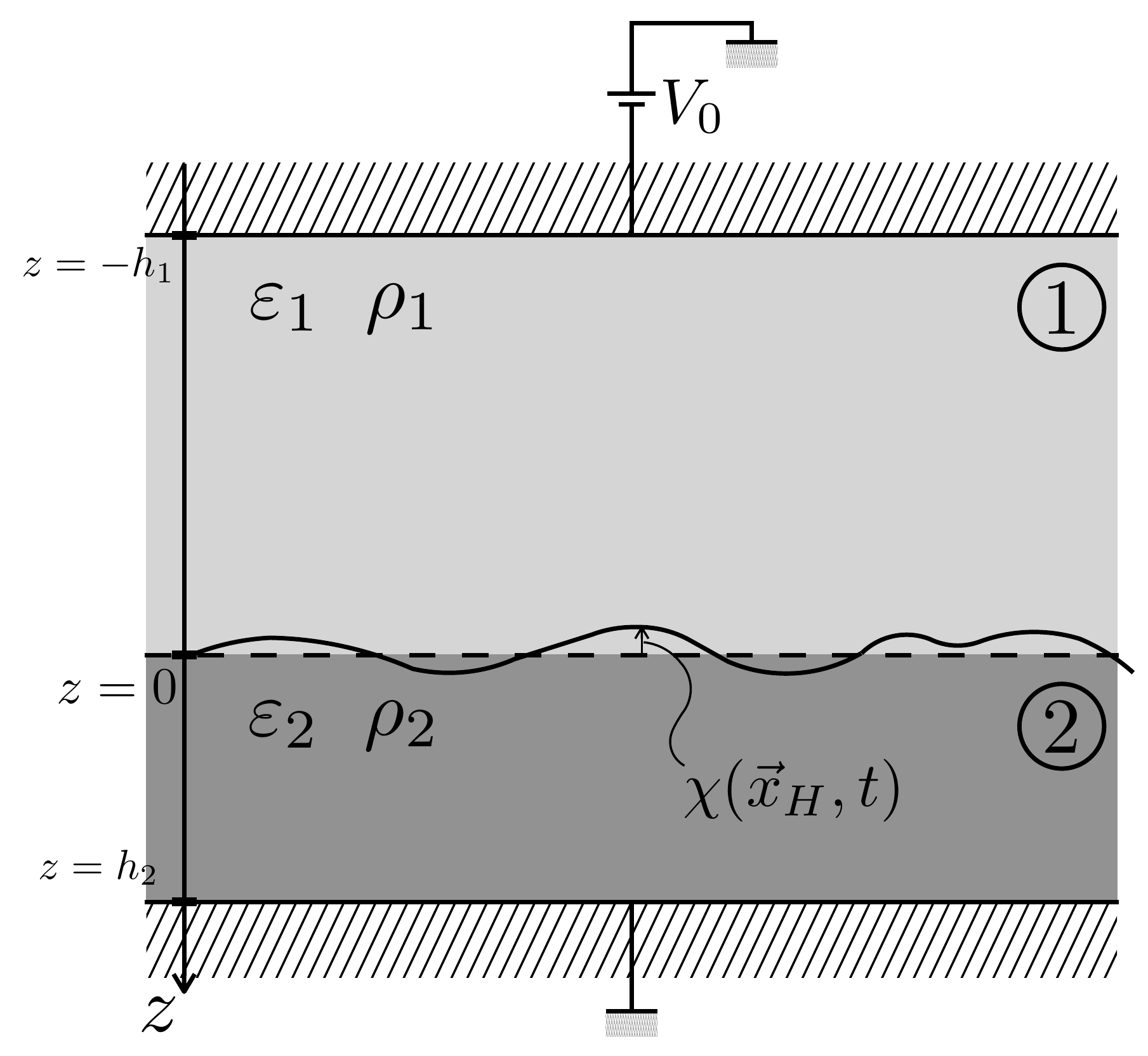}%
 \caption{Schematic representation of a system section showing the two immiscible fluids and the interface between them $\chi(x,y,t)$.}
\label{systemschematics}
\end{figure}

Let us consider two horizontal liquid layers in a laterally bounded domain and confined vertically between two horizontal parallel electrodes. The liquids are assumed to be perfect dielectrics with electric permittivities $\varepsilon_1$ and $\varepsilon_2$. Subindex 1 corresponds to the upper layer, and subindex 2 to the lower one. We also suppose that they are ideal liquids, i.e., we neglect viscosity. Their densities are denoted $\rho_1$ and $\rho_2$. In the unperturbed state, the upper layer thickness is denoted $h_1$ and the lower one $h_2$. The upper electrode is at a voltage $V_0$ and the lower one is grounded. The coordinate system is chosen in such a way that the $z$ coordinate is pointing downwards, and $z=0$ corresponds to the unperturbed interface. We suppose that the interface is perturbed in such a way that it is  defined by the function $z=\chi(x,y,t)$. This description is illustrated in Fig. \ref{systemschematics}.

The system is subjected to vertical oscillations of amplitude $\epsilon$ and frequency $\omega$. To describe the effect of this oscillations we consider Cartesian axes $(x,y,z)$ moving with the vessel. The motion relative to these axes is the same as the one due to a gravitational acceleration given by $g-\epsilon\cos(\omega t)$. The following treatment is based on \cite{Benjamin1954}. 

In each region the continuity equation and the equation of motion,
\begin{equation}
\nabla\cdot\mathbf{v}_i=0, \quad\quad \rho_i\frac{\partial \mathbf{v}_i}{\partial t}+\nabla p_i=\rho_i(g-\epsilon\cos(\omega t))\mathbf{e}_z, \ (i=1,2) \label{bulk}
\end{equation}
are satisfied, where $\mathbf{v}_i$ is the liquid velocity, $p_i$ the pressure, and $\mathbf{e}_z$ is the unitary vector in $z$-direction. 

Neglecting viscosity allows to write the liquid velocity as the gradient of a potential, $\mathbf{v}_i=\nabla\psi_i$. Equations in (\ref{bulk}) are expressed in terms of  $\psi$ as
\begin{eqnarray}
\nabla^2\psi_i&=&0 \label{euler_laplace},\\
\rho_i\frac{\partial \nabla\psi_i}{\partial t}+\nabla p_i&=&\rho_i(g-\epsilon\cos(\omega t))\mathbf{e}_z \label{euler_potential}. 
\end{eqnarray} 

Added to the hydrodynamic side of the problem, in each region the electric potential $\Phi$ is obtained from Laplace's equation
\begin{equation}
\nabla^2\Phi_i=0. \quad (i=1,2) \label{laplace}
\end{equation}

We have to compliment these equations with the corresponding boundary conditions. The fact that the liquid surface is a material one is expressed by the kinematic condition
\begin{equation}
-\frac{\partial\chi}{\partial t}+v_z-v_x\frac{\partial\chi}{\partial x}-v_y\frac{\partial\chi}{\partial y}=0. \label{kinematic}
\end{equation}
The balance of normal stresses at the interface is given by
\begin{equation}
-[p]+\mathbf{n}\cdot[\overline{\overline T}_M]\cdot\mathbf{n}-\gamma\nabla_s\cdot\mathbf{n}=0, \label{normal_balance}
\end{equation}
where $\mathbf{n}$ is a vector normal to the volume element, $\overline{\overline T}_M$ is Maxwell's stress tensor, $\gamma$ is the surface tension coefficient, and $\nabla_s$ is the surface gradient operator. The brackets of any quantity $[A]$ denotes its jump across the interface: $[A]\equiv A_2-A_1$.
Maxwell's stress tensor, neglecting electrostriction, is given by
\begin{equation}
T_{Mij}=\varepsilon\left(E_iE_j-\frac12\delta_{ij}E^2\right).
\end{equation}

The mechanical boundary conditions at the electrodes are
\begin{equation}
v_z=0 \quad \mbox{at}\quad z=-h_1,h_2. \label{velocity_electrodes}
\end{equation}

The electrical boundary conditions are
\begin{eqnarray}
[\mathbf{E}]\times \mathbf{n}=0 && \quad\mbox{at the interface,} \label{et_interface}\\ 
\mbox{[}\varepsilon\mathbf{E}\mbox{]}\cdot\mathbf{n} =0 && \quad \mbox{at the interface,} \label{en_interface}\\
\Phi_1=V_0 && \quad\mbox{at} \quad z=-h_1, \\
\Phi_2=0 && \quad \mbox{at} \quad z=h_2.
\end{eqnarray}
The first of these conditions is equivalent to the continuity of the electric potential through the interface. The second one represents the absence of free surface charge at the interface.

The balance of stresses at the interface deserves a comment. The normal stress is always continuous through a liquid interface. However, this is not the case for the tangential stresses. If viscosity is neglected, requiring the continuity of tangential stresses will overdetermine the problem. Neglecting viscosity allows a discontinuity of the tangential velocity component  (see \cite{Melcher1969,Castellanos1998} for a detailed discussion).


\subsection{Solution for an unperturbed interface}

For a planar interface $\chi=0,$ both liquids are at rest, i.e., $\mathbf{v}=0$. The pressure in each fluid layer is
\begin{equation}
p_{0i}=\Pi_{0i}+\rho_i(g-\epsilon\cos(\omega t))z. \quad (i=1,2)
\end{equation}
The electric pressure entails a pressure difference given by equation (\ref{normal_balance}),
\begin{equation}
\Pi_{02}-\Pi_{01}=\frac12\varepsilon_2E_2^2-\frac12\varepsilon_1E_1^2. \label{normal_balance_zero}
\end{equation}
 
In the unperturbed state, the electric potential $\Phi_0(z)$ is a linear function of $z$ in each region. Applying boundary conditions  $\Phi_0(-h_1)=V_0$ and $\Phi_0(h_2)=0$ yields
\begin{eqnarray}
\Phi_{01}(z)&=&V_0\left(1-\frac{\varepsilon_2}{\varepsilon_1h_2+\varepsilon_2h_1}(z+h_1)\right),\\
\Phi_{02}(z)&=&V_0\frac{\varepsilon_1}{\varepsilon_1h_2+\varepsilon_2h_1}(h_2-z),
\end{eqnarray}
so that electric field in each region becomes
\begin{eqnarray}
\mathbf{E}_{01}(z)&=&\frac{\varepsilon_2}{\varepsilon_1h_2+\varepsilon_2h_1}V_0\mathbf{e}_z,\\
\mathbf{E}_{02}(z)&=&\frac{\varepsilon_1}{\varepsilon_1h_2+\varepsilon_2h_1}V_0\mathbf{e}_z.
\end{eqnarray}

\subsection{Linear perturbation}

Let us assume that the surface is perturbed in such a way that $\chi(x,y)\ll h_1,h_2$. It is then possible to expand all functions as a power series, retaining only linear terms. Special care must be taken to expand the values of any function at the interface. The general procedure  for any quantity $A(x,y,z)$ is \cite{Melcher1963,Lighthill1978}
\begin{equation}
A(x,y,z=\chi)=A_0(x,y,z=0)+\left.\frac{\partial A_0}{\partial z}\right|_{z=0}\chi+\delta A(x,y,z=0)+\ldots \label{expansion}
\end{equation}

To first order, equation (\ref{euler_potential}) becomes
\begin{equation}
\rho_i\frac{\partial \psi_i}{\partial t}+\delta p_i =0, \quad (i=1,2) \label{pressure_potential}
\end{equation}
where $\psi_i$ is the velocity potential, and where the constant of integration is taken to be zero.

Let us consider the three terms of equation (\ref{normal_balance}). The first one is [see equation (\ref{expansion})]
\begin{eqnarray}
[p]=p_{02}(z=0)&+&\frac{\partial p_{02}}{\partial z}\chi+\delta p_2(0) \nonumber\\ 
-p_{01}(z=0)&-&\frac{\partial p_{01}}{\partial z}\chi-\delta p_1(0)\nonumber\\
&=& \Pi_{02}-\Pi_{01}+(\rho_2-\rho_1)(g-\epsilon\cos(\omega t))\chi+\delta p_2(0)-\delta p_1(0). \label{pressure_int}
\end{eqnarray}

For the electric term we have
\begin{equation}
\mathbf{n}\cdot\overline{\overline T}_{Mi}\cdot\mathbf{n}=T_{Mizz}=
	\frac12\varepsilon_iE_{zi}^2=\frac12\varepsilon_iE_{0i}^2-\varepsilon_iE_{0i}\left.\frac{\partial\delta \Phi_i}{\partial z}\right|_{z=0}\chi, \label{electric_int}
\end{equation}
where all  terms of order higher than one have been neglected. Here, we have taken into account that, to the linear approximation, the unit vector normal to the interface is given by 
\begin{equation}
\mathbf{n}=-\frac{\partial\chi}{\partial x}\mathbf{e}_x-\frac{\partial\chi}{\partial y}\mathbf{e}_y+\mathbf{e}_z.
\end{equation}

The last term in  equation (\ref{normal_balance}) becomes
\begin{equation}
-\gamma\nabla_s\cdot\mathbf{n}=\gamma\nabla_s^2\chi. \label{tension_in}
\end{equation}

Collecting (\ref{pressure_int}), (\ref{electric_int}) and (\ref{tension_in}), the first order approximation to the stress normal balance then becomes
\begin{eqnarray}
-\delta p_2(z=0)+\delta p_1(z=0)&-&(\rho_2-\rho_1)(g-\epsilon\cos(\omega t))\chi, \\
-\varepsilon_2E_{02}\left.\frac{\partial\delta \Phi_2}{\partial z}\right|_{z=0}&+
&\varepsilon_1E_{01}\left.\frac{\partial\delta \Phi_1}{\partial z}\right|_{z=0}+\gamma\nabla_s^2\chi=0,
\end{eqnarray}
where equation (\ref{normal_balance_zero}) has been used to remove the static pressure.

With the help of (\ref{pressure_potential}), last equation can be written as
\begin{equation}
\rho_2\left.\frac{\partial \psi_2}{\partial t}\right|_{z=0}-\rho_1\left.\frac{\partial \psi_1}{\partial t}\right|_{z=0}-(\rho_2-\rho_1)(g-\epsilon\cos(\omega t))\chi
-\varepsilon_2E_{02}\left.\frac{\partial\delta \Phi_2}{\partial z}\right|_{z=0}+
\varepsilon_1E_{01}\left.\frac{\partial\delta \Phi_1}{\partial z}\right|_{z=0}+\gamma\nabla_s^2\chi=0. \label{interface_linear}
\end{equation}

The linear approximation of equation (\ref{kinematic}) relates $\psi$ and $\chi$,
\begin{equation}
\frac{\partial\chi}{\partial t}=\frac{\partial\psi}{\partial z} \quad \mbox{at} \quad z=0. \label{linear_kinematic}
\end{equation}

Following Benjamin and Ursell \cite{Benjamin1954}, the function $\chi(x,y,t)$ can be expanded in terms of a complete orthogonal set of eigenfunctions $S_m(x,y)$ that fulfill
\begin{equation}
\left(\frac{\partial^2}{\partial x^2}+\frac{\partial^2}{\partial y^2}+k_m^2\right)S_m(x,y)=0
\end{equation}
with adequate boundary conditions on the lateral walls of the domain (an example of the use of this type of expansion for anular geometry may be found in \cite{Gonzalez1997}). Therefore, taking $\chi(x,y,t)=\sum_{m} a_m(t)S_m(x,y)$, the solution of (\ref{euler_laplace}) that obeys the boundary conditions (\ref{velocity_electrodes}) and (\ref{linear_kinematic}) is
\begin{eqnarray}
\psi_1(x,y,z,t)&=&\sum_{m}\frac{1}{k_m\sinh(k_mh_1)}\frac{da_m(t)}{dt}S_m(x,y)\cosh(k_m(z+h_1)) \label{potential_velocity_1},\\
\psi_2(x,y,z,t)&=&-\sum_{m}\frac{1}{k_m\sinh(k_mh_2)}\frac{da_m(t)}{dt}S_m(x,y)\cosh(k_m(z-h_2)) \label{potential_velocity_2}.
\end{eqnarray}

The case of the electric potential is a bit trickier. Equation (\ref{en_interface}) at first order becomes
\begin{equation}
\varepsilon_1\left.\frac{\partial\delta\Phi_1}{\partial z}\right|_{z=0}=\varepsilon_2\left.\frac{\partial\delta\Phi_2}{\partial z}\right|_{z=0}. \label{en_interface_linear}
\end{equation}
 On the other hand, equation (\ref{et_interface}) is equivalent to the continuity of the potential through the interface. Taking into account (\ref{expansion}), it is
 \begin{equation}
 \delta\Phi_1(0)-E_{01}\chi=\delta\Phi_2(0)-E_{02}\chi. \label{et_interface_linear}
 \end{equation}
 
 The solution of Laplace's equation that fulfil (\ref{en_interface_linear})-(\ref{et_interface_linear}) and $\delta \Phi=0$ at $z=-h_1,h_2$ is
 \begin{eqnarray}
 \delta\Phi_1(x,y,z,t)&=&\sum_{m}\frac{\varepsilon_2}{\cosh(k_mh_1)}\frac{(\varepsilon_2-\varepsilon_1)V_0}{\varepsilon_{\mathrm{eq},m}(\varepsilon_2h_1+\varepsilon_1h_2)}a_m(t)S_m(x,y)\sinh(k_m(z+h_1)), \label{potential_field_1}\\
 \delta\Phi_2(x,y,z,t)&=&\sum_{m}\frac{\varepsilon_1}{\cosh(k_mh_2)}\frac{(\varepsilon_2-\varepsilon_1)V_0}{\varepsilon_{\mathrm{eq},m}(\varepsilon_2h_1+\varepsilon_1h_2)}a_m(t)S_m(x,y)\sinh(k_m(z-h_2)), \label{potential_field_2}
\end{eqnarray}
where the factor $\varepsilon_{\mathrm{eq},m}$ is given by $\varepsilon_{\mathrm{eq},m}=\varepsilon_2\tanh(k_mh_1)+\varepsilon_1\tanh(k_mh_2)$.

Introducing (\ref{potential_velocity_1},\ref{potential_velocity_2},\ref{potential_field_1},\ref{potential_field_2}) into (\ref{interface_linear}), and making use of the orthogonality of functions $S_m(x,y)$ we arrive at
\begin{equation}
\frac{d^2a_m(t)}{dt^2}+\left[\frac{\gamma k_m^3}{\rho_{eq,m}}+\frac{(g-\epsilon\cos(\omega t))k_m(\rho_2-\rho_1)}{\rho_{\mathrm{eq},m}}-\frac{\varepsilon_1\varepsilon_2 k_m^2}{\varepsilon_{\mathrm{eq},m}\rho_{\mathrm{eq},m}}\left(\frac{V_0(\varepsilon_1-\varepsilon_2)^2}{\varepsilon_2h_1+\varepsilon_1h_2}\right)^2\right]a_m(t)=0 \label{equation_for_a_m},
\end{equation}
with $\rho_{\mathrm{eq},m}=\rho_1\coth(k_mh_1)+\rho_2\coth(k_mh_2)$.

In the absence of parametric forcing, and for a continuous $k$ spectrum, this equation indeed coincides with the one obtained by Melcher for Electrohydrodynamic waves between two dielectric liquids in the absence of free charge at the interface \cite{Melcher1961,Melcher1963}.

%
\section{Mathieu's equation}

Equation (\ref{equation_for_a_m}) has the form of a Mathieu's equation,
\begin{equation}
 \frac{d^2a_m}{dT^2}+(p_m-2q_m\cos(2T))a_m = 0, \label{mathieu} 
\end{equation}
where the coefficients are defined by
$$p_m \equiv\left[\frac{4\gamma k_m^3}{\omega^2\rho_{eq,m}}+\frac{4gk_m(\rho_2-\rho_1)}{\omega^2\rho_{eq,m}}-\frac{4\varepsilon_1\varepsilon_2 k_m^2}{\omega^2f_m\rho_{eq,m}}\left(\frac{V_0(\varepsilon_1-\varepsilon_2)^2}{\varepsilon_2h_1+\varepsilon_1h_2}\right)^2\right],$$
\begin{equation}
q_m \equiv \frac{2k_m\epsilon(\rho_2-\rho_1)}{\omega^2\rho_{eq,m}}, \quad T \equiv \frac{1}{2}\omega t. \label{coefficients} 
\end{equation}

This equation is similar to the one obtained by Benjamin and Ursell for the Faraday waves on a liquid surface, although Benjamin and Ursell treatment considers only one liquid layer.  The novelty here is the appearance of the electric term that modifies the coeffecient $p_m$. The equation is also similar to those obtained by Briskman and Shaidurov \cite{Briskman1968} and Yih \cite{Yih1968}, or Gonz\'alez et al.  \cite{Gonzalez1997}. The difference relies in that in those cases the electric field modifies the coefficient $q_m$, whereas in our case is the coefficient $p_m$ that becomes affected.

\begin{figure}
\includegraphics[width=14cm]{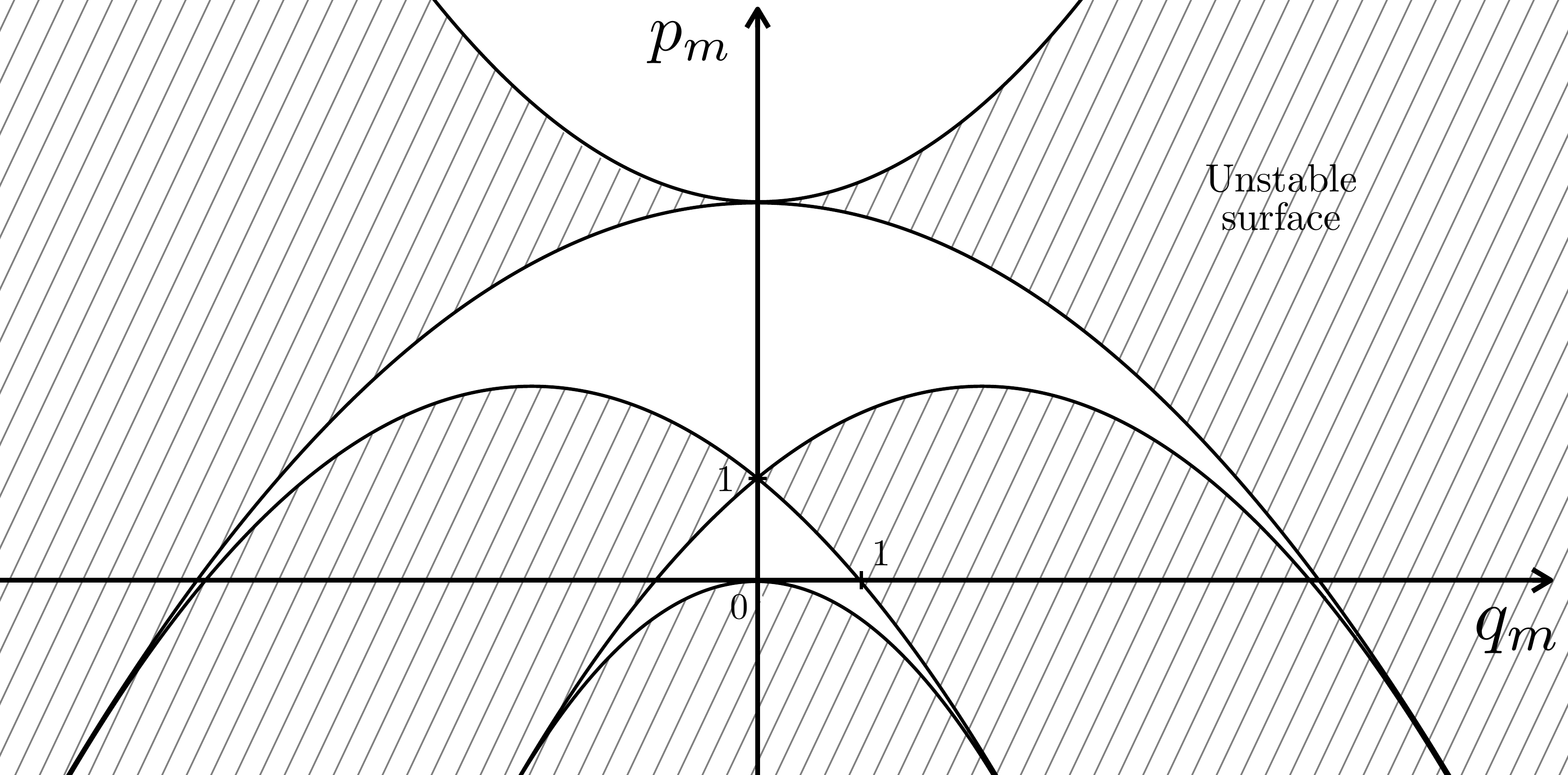}%
 \caption{Mathieu's instability chart. Shadowed regions correspond to unstable solutions, while white ones are stable. It can be observed that in doing a vertical movement in this chart one may achieve both effects: a stable surface to become unstable, and an unstable surface to become unstable, as discussed in last section.}
\label{mathieuchart}
\end{figure}

Arranging the equation in the form of a Matieu equation (\ref{mathieu}) has the great benefice of the already existing stability analysis of this equation. We point out the one shown in \cite{Bender1999} because of the negative values it presents, as the addition of the electric field now allows a negative contribution to the $p_m$ coefficients. This chart is exposed here schematically in Fig. \ref{mathieuchart}.

\section{Discussion and conclusion} \label{conclusion}

Taking a closer look at the shape of the chart shown in Fig. \ref{mathieuchart}, it can be easily seen that for any given configuration, it always exists a potential difference between the two electrodes that can destabilise the surface, causing the appearance of  standing waves.

\begin{figure}
\includegraphics[width=9cm]{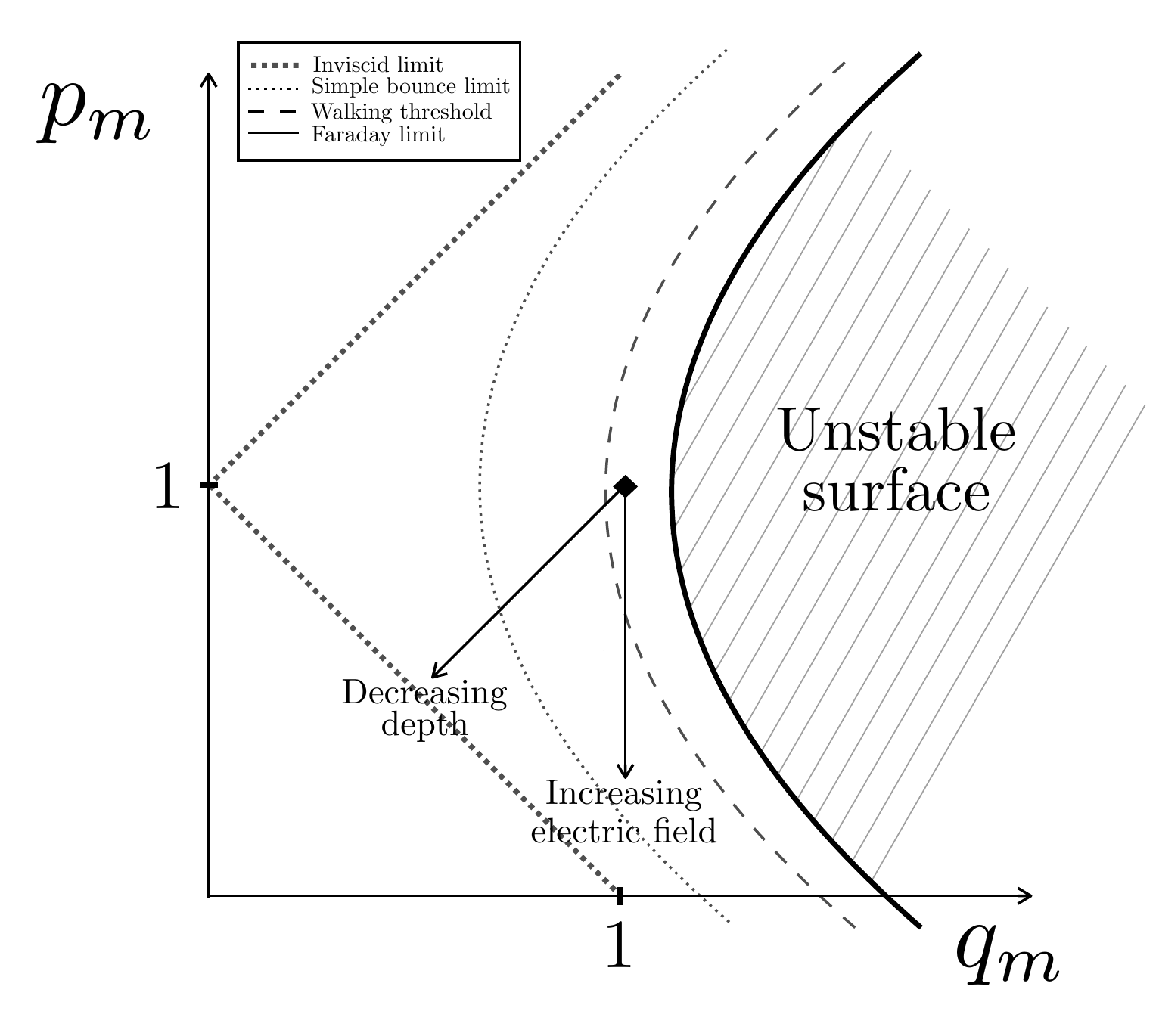}%
 \caption{Schematic effect of the modification of the operation point if the depth is modified or an electric potential is placed, given (\ref{mathieu}). It is here represented the first peak in the inviscid limit studied in this work, and a qualitative raising of it in the presence of viscosity to show a more realistic scenery for the case of the walking droplets, described in the image.}
\label{electriceffect}
\end{figure}

Another noteworthy application for the inclusion of an electric field is a reverse one from the above. Given the successive peaks in the solution chart of the equation, one might have a configuration in a way that the operation point is unstable if an electric field is not present, but can be turned stable when a potential difference between the two electrodes is placed. The Mathieu chart is modified once viscosity is included in the analysis.  It can be shown that its main effect is to pull apart the minimum values from $q=0$ to finite ones \citep{Kumar1996}. An schematic view is indicated in Fig. \ref{electriceffect}. 
In this figure we have plot the inviscid limit of stability and the stability threshold when viscosity is taken into account. In the shadowed region the surface becomes unstable and Faraday waves are produced. The phenomenon of bouncing drops occurs near the instability threshold, but in the stable region, where Faraday waves decay in time. The limit of walking droplets is also plotted. 

The current strategy to confine the walking droplets is through the container depth. In Fig. \ref{electriceffect} we have indicated the effect of changing the depth in the stability chart. This effect is due to $\rho_{\mathrm{eq},m}$, which affects both $q_m$ and $p_m$. Decreasing the depth takes the system out of the walking region, or even from the existence of bouncing droplets. The practical consequence is to create an effective wall for the droplets. A similar effect can be obtained increasing the electric field, although is a bit less effective, due to the fact that the electric field only modifies the coefficient $p_m$. However, the electric field is more versatile from the experimental point of view. The field can be switch on and off, thus allowing the drop to cross or not a given border at will.

The value of the field which is needed to achieve the unstability for any given configuration in two realistic cases is obtained, one taken from \cite{Protiere2006} and the other from \cite{Molacek2013}. In the first one, a fluid of $\gamma = 0.0209\ \text{N\ m}^{-1}$ and $\rho = 965\ \text{Kg\ m}^{-3}$ is confined in a square container $(8\times8\times1$ cm), partially filling it ($h=1$ mm). For the second one, a liquid of  $\gamma = 0.0208\ \text{N\ m}^{-1}$ and $\rho = 949\ \text{Kg\ m}^{-3}$ is placed in a circular recipient of 76 mm of diameter, filling it to a liquid depth of $h=9$ mm.

Taking into account that in both of these cases the fluids are exposed to free air so that in (\ref{coefficients}) $\rho_2\gg\rho_1$, for the first case we obtain a value of 20 V for the (1,1) mode, and for the second one 10 V for the lower (1,0) mode. It is therefore shown that with relatively low fields (in comparison with the dielectric rupture potentials which stands at around $0.4\times10^4$ V for the case of air when a separation between electrodes is considered to be 1 cm, as in the case of \cite{Protiere2006}) confinement of droplets can be obtained. From an experimental point of view the use of an AC electric field is recommended \cite{Gonzalez1989}. The AC electric field avoids the possible influence of Coulomb forces. However, the frequency of the electric field must be much greater than any relevant mechanical frequency.

In conclusion, we have obtained the dispersion relation for Electrohydrodynamic waves on an oscillating liquid surface. The obtained results open the possibility of using the electric field to control walking droplets on oscillating liquid surfaces.

\begin{acknowledgments}
	This work was funded by  the Spanish Ministerio de Ciencia, Innovación y Universidades  under Research Project No.PGC2018-099217-B-I00, and Junta de Andalucía under research project 2017/FQM-253.
\end{acknowledgments}


\begin{thebibliography}{21}
	\makeatletter
	\providecommand \@ifxundefined [1]{%
		\@ifx{#1\undefined}
	}%
	\providecommand \@ifnum [1]{%
		\ifnum #1\expandafter \@firstoftwo
		\else \expandafter \@secondoftwo
		\fi
	}%
	\providecommand \@ifx [1]{%
		\ifx #1\expandafter \@firstoftwo
		\else \expandafter \@secondoftwo
		\fi
	}%
	\providecommand \natexlab [1]{#1}%
	\providecommand \enquote  [1]{``#1''}%
	\providecommand \bibnamefont  [1]{#1}%
	\providecommand \bibfnamefont [1]{#1}%
	\providecommand \citenamefont [1]{#1}%
	\providecommand \href@noop [0]{\@secondoftwo}%
	\providecommand \href [0]{\begingroup \@sanitize@url \@href}%
	\providecommand \@href[1]{\@@startlink{#1}\@@href}%
	\providecommand \@@href[1]{\endgroup#1\@@endlink}%
	\providecommand \@sanitize@url [0]{\catcode `\\12\catcode `\$12\catcode
		`\&12\catcode `\#12\catcode `\^12\catcode `\_12\catcode `\%12\relax}%
	\providecommand \@@startlink[1]{}%
	\providecommand \@@endlink[0]{}%
	\providecommand \url  [0]{\begingroup\@sanitize@url \@url }%
	\providecommand \@url [1]{\endgroup\@href {#1}{\urlprefix }}%
	\providecommand \urlprefix  [0]{URL }%
	\providecommand \Eprint [0]{\href }%
	\providecommand \doibase [0]{https://doi.org/}%
	\providecommand \selectlanguage [0]{\@gobble}%
	\providecommand \bibinfo  [0]{\@secondoftwo}%
	\providecommand \bibfield  [0]{\@secondoftwo}%
	\providecommand \translation [1]{[#1]}%
	\providecommand \BibitemOpen [0]{}%
	\providecommand \bibitemStop [0]{}%
	\providecommand \bibitemNoStop [0]{.\EOS\space}%
	\providecommand \EOS [0]{\spacefactor3000\relax}%
	\providecommand \BibitemShut  [1]{\csname bibitem#1\endcsname}%
	\let\auto@bib@innerbib\@empty
\bibitem [{\citenamefont {Melcher}(1961)}]{Melcher1961}%
\BibitemOpen
\bibfield  {author} {\bibinfo {author} {\bibfnamefont {J.~R.}\ \bibnamefont
		{Melcher}},\ }\bibfield  {title} {\bibinfo {title} {Electrohydrodynamic and
		magnetohydrodynamic surface waves and instabilites},\ }\href@noop {}
{\bibfield  {journal} {\bibinfo  {journal} {Phys. of Fluids}\ }\textbf
	{\bibinfo {volume} {4}},\ \bibinfo {pages} {1348} (\bibinfo {year}
	{1961})}\BibitemShut {NoStop}%
\bibitem [{\citenamefont {Melcher}(1963)}]{Melcher1963}%
\BibitemOpen
\bibfield  {author} {\bibinfo {author} {\bibfnamefont {J.~R.}\ \bibnamefont
		{Melcher}},\ }\href@noop {} {\emph {\bibinfo {title} {Field-Coupled Surface
			Waves. A comparative study of surface-coupled electrohydrodynamic and
			magnetohydrodynamic systems}}}\ (\bibinfo  {publisher} {The M.I.T. Press},\
\bibinfo {year} {1963})\BibitemShut {NoStop}%
\bibitem [{\citenamefont {Taylor}\ and\ \citenamefont
	{McEwan}(1965)}]{Taylor1965}%
\BibitemOpen
\bibfield  {author} {\bibinfo {author} {\bibfnamefont {G.~I.}\ \bibnamefont
		{Taylor}}\ and\ \bibinfo {author} {\bibfnamefont {A.~D.}\ \bibnamefont
		{McEwan}},\ }\bibfield  {title} {\bibinfo {title} {The stability of a
		horizontal fluid interface in a vertical electric field},\ }\href@noop {}
{\bibfield  {journal} {\bibinfo  {journal} {J. Fluid Mech.}\ }\textbf
	{\bibinfo {volume} {22}},\ \bibinfo {pages} {1} (\bibinfo {year}
	{1965})}\BibitemShut {NoStop}%
\bibitem [{\citenamefont {Melcher}\ and\ \citenamefont
	{Smith~Jr.}(1969)}]{Melcher1969}%
\BibitemOpen
\bibfield  {author} {\bibinfo {author} {\bibfnamefont {J.~R.}\ \bibnamefont
		{Melcher}}\ and\ \bibinfo {author} {\bibfnamefont {C.~W.}\ \bibnamefont
		{Smith~Jr.}},\ }\bibfield  {title} {\bibinfo {title} {Electrohydrodynamic
		charge relaxation and interfacial perpendicular-field instability},\
}\href@noop {} {\bibfield  {journal} {\bibinfo  {journal} {Phys. of Fluids}\
	}\textbf {\bibinfo {volume} {12}},\ \bibinfo {pages} {778} (\bibinfo {year}
	{1969})}\BibitemShut {NoStop}%
\bibitem [{\citenamefont {Kath}\ and\ \citenamefont {Hoburg}(1977)}]{Kath1977}%
\BibitemOpen
\bibfield  {author} {\bibinfo {author} {\bibfnamefont {G.~S.}\ \bibnamefont
		{Kath}}\ and\ \bibinfo {author} {\bibfnamefont {J.~F.}\ \bibnamefont
		{Hoburg}},\ }\bibfield  {title} {\bibinfo {title} {Interfacial
		electrohydrodynamic instability in normal electric field},\ }\href
{https://doi.org/10.1063/1.861978} {\bibfield  {journal} {\bibinfo  {journal}
		{The Physics of Fluids}\ }\textbf {\bibinfo {volume} {20}},\ \bibinfo {pages}
	{912} (\bibinfo {year} {1977})},\ \Eprint
{https://arxiv.org/abs/https://aip.scitation.org/doi/pdf/10.1063/1.861978}
{https://aip.scitation.org/doi/pdf/10.1063/1.861978} \BibitemShut {NoStop}%
\bibitem [{\citenamefont {Chu}\ \emph {et~al.}(1989)\citenamefont {Chu},
	\citenamefont {Velarde},\ and\ \citenamefont {Castellanos}}]{Chu1989}%
\BibitemOpen
\bibfield  {author} {\bibinfo {author} {\bibfnamefont {X.~L.}\ \bibnamefont
		{Chu}}, \bibinfo {author} {\bibfnamefont {M.~G.}\ \bibnamefont {Velarde}},\
	and\ \bibinfo {author} {\bibfnamefont {A.}~\bibnamefont {Castellanos}},\
}\bibfield  {title} {\bibinfo {title} {Dissipative hydrodynamic oscillators.
		iii. - electrohydrodynamic interfacial waves},\ }\href@noop {} {\bibfield
	{journal} {\bibinfo  {journal} {Il Nuovo Cimento}\ }\textbf {\bibinfo
		{volume} {11D}},\ \bibinfo {pages} {727} (\bibinfo {year}
	{1989})}\BibitemShut {NoStop}%
\bibitem [{\citenamefont {Castellanos}\ and\ \citenamefont
	{Gonz{\'a}lez}(1992)}]{Castellanos1992}%
\BibitemOpen
\bibfield  {author} {\bibinfo {author} {\bibfnamefont {A.}~\bibnamefont
		{Castellanos}}\ and\ \bibinfo {author} {\bibfnamefont {A.}~\bibnamefont
		{Gonz{\'a}lez}},\ }\bibfield  {title} {\bibinfo {title} {Interfacial
		electrohydrodynamic instability: The kath and hoburg model revisited},\
}\href@noop {} {\bibfield  {journal} {\bibinfo  {journal} {Phys. of Fluids}\
	}\textbf {\bibinfo {volume} {4}},\ \bibinfo {pages} {1307} (\bibinfo {year}
	{1992})}\BibitemShut {NoStop}%
\bibitem [{\citenamefont {Castellanos}\ and\ \citenamefont
	{Gonz{\'a}lez}(1998)}]{Castellanos1998}%
\BibitemOpen
\bibfield  {author} {\bibinfo {author} {\bibfnamefont {A.}~\bibnamefont
		{Castellanos}}\ and\ \bibinfo {author} {\bibfnamefont {A.}~\bibnamefont
		{Gonz{\'a}lez}},\ }\bibfield  {title} {\bibinfo {title} {Nonlinear
		electrohydrodynamics of free surfaces},\ }\href@noop {} {\bibfield  {journal}
	{\bibinfo  {journal} {IEE Transactions on Dielectrics and Electrical
			Insulation}\ }\textbf {\bibinfo {volume} {5}},\ \bibinfo {pages} {334}
	(\bibinfo {year} {1998})}\BibitemShut {NoStop}%
\bibitem [{\citenamefont {Briskman}\ and\ \citenamefont
	{Shaidurov}(1968)}]{Briskman1968}%
\BibitemOpen
\bibfield  {author} {\bibinfo {author} {\bibfnamefont {V.~A.}\ \bibnamefont
		{Briskman}}\ and\ \bibinfo {author} {\bibfnamefont {G.~F.}\ \bibnamefont
		{Shaidurov}},\ }\bibfield  {title} {\bibinfo {title} {Parametric instability
		of a fluid surface in an alternating electric field},\ }\href@noop {}
{\bibfield  {journal} {\bibinfo  {journal} {Dokl. Akad. Nauk SSSR}\ }\textbf
	{\bibinfo {volume} {180}},\ \bibinfo {pages} {1315} (\bibinfo {year}
	{1968})}\BibitemShut {NoStop}%
\bibitem [{\citenamefont {Yih}(1968)}]{Yih1968}%
\BibitemOpen
\bibfield  {author} {\bibinfo {author} {\bibfnamefont {C.~S.}\ \bibnamefont
		{Yih}},\ }\bibfield  {title} {\bibinfo {title} {Stability of a horizontal
		fluid interface in a periodic vertical electric field},\ }\href@noop {}
{\bibfield  {journal} {\bibinfo  {journal} {Phys. of Fluids}\ }\textbf
	{\bibinfo {volume} {11}},\ \bibinfo {pages} {1447} (\bibinfo {year}
	{1968})}\BibitemShut {NoStop}%
\bibitem [{\citenamefont {Gonz{\'a}lez}\ \emph {et~al.}(1997)\citenamefont
	{Gonz{\'a}lez}, \citenamefont {Ramos},\ and\ \citenamefont
	{Castellanos}}]{Gonzalez1997}%
\BibitemOpen
\bibfield  {author} {\bibinfo {author} {\bibfnamefont {A.}~\bibnamefont
		{Gonz{\'a}lez}}, \bibinfo {author} {\bibfnamefont {A.}~\bibnamefont
		{Ramos}},\ and\ \bibinfo {author} {\bibfnamefont {A.}~\bibnamefont
		{Castellanos}},\ }\bibfield  {title} {\bibinfo {title} {Parametric
		instability of dielectric, slightly viscous liquid jets under ac electric
		fields},\ }\href@noop {} {\bibfield  {journal} {\bibinfo  {journal} {Phys. of
			Fluids}\ }\textbf {\bibinfo {volume} {9}},\ \bibinfo {pages} {1830} (\bibinfo
	{year} {1997})}\BibitemShut {NoStop}%
\bibitem [{\citenamefont {Ward}\ \emph {et~al.}(2019)\citenamefont {Ward},
	\citenamefont {Matsumoto},\ and\ \citenamefont {Narayanan}}]{Ward2019}%
\BibitemOpen
\bibfield  {author} {\bibinfo {author} {\bibfnamefont {K.}~\bibnamefont
		{Ward}}, \bibinfo {author} {\bibfnamefont {S.}~\bibnamefont {Matsumoto}},\
	and\ \bibinfo {author} {\bibfnamefont {R.}~\bibnamefont {Narayanan}},\
}\bibfield  {title} {\bibinfo {title} {The electrostatically forced faraday
		instability: theory and experiments},\ }\href@noop {} {\bibfield  {journal}
	{\bibinfo  {journal} {J. Fluid Mech.}\ }\textbf {\bibinfo {volume} {862}},\
	\bibinfo {pages} {696} (\bibinfo {year} {2019})}\BibitemShut {NoStop}%
\bibitem [{\citenamefont {Benjamin}\ and\ \citenamefont
	{Ursell}(1954)}]{Benjamin1954}%
\BibitemOpen
\bibfield  {author} {\bibinfo {author} {\bibfnamefont {T.}~\bibnamefont
		{Benjamin}}\ and\ \bibinfo {author} {\bibfnamefont {F.}~\bibnamefont
		{Ursell}},\ }\bibfield  {title} {\bibinfo {title} {The stability of the plane
		free surface of a liquid in vertical periodic motion},\ }\href@noop {}
{\bibfield  {journal} {\bibinfo  {journal} {Proc. R. Soc.}\ }\textbf
	{\bibinfo {volume} {225}},\ \bibinfo {pages} {505} (\bibinfo {year}
	{1954})}\BibitemShut {NoStop}%
\bibitem [{\citenamefont {McLachlan}(1947)}]{Mclachlan1947}%
\BibitemOpen
\bibfield  {author} {\bibinfo {author} {\bibfnamefont {N.~W.}\ \bibnamefont
		{McLachlan}},\ }\href@noop {} {\emph {\bibinfo {title} {Theory and
			application of Mathieu functions}}}\ (\bibinfo  {publisher} {Oxford
	University Press},\ \bibinfo {year} {1947})\BibitemShut {NoStop}%
\bibitem [{\citenamefont {Kumar}(1996)}]{Kumar1996}%
\BibitemOpen
\bibfield  {author} {\bibinfo {author} {\bibfnamefont {K.}~\bibnamefont
		{Kumar}},\ }\bibfield  {title} {\bibinfo {title} {Linear theory of faraday
		instability in viscous liquids},\ }\href@noop {} {\bibfield  {journal}
	{\bibinfo  {journal} {Proc. R. Soc. Lond. A}\ }\textbf {\bibinfo {volume}
		{452}},\ \bibinfo {pages} {1113} (\bibinfo {year} {1996})}\BibitemShut
{NoStop}%
\bibitem [{\citenamefont {Proti{\`e}re}\ \emph {et~al.}(2006)\citenamefont
	{Proti{\`e}re}, \citenamefont {Boudaoud},\ and\ \citenamefont
	{Couder}}]{Protiere2006}%
\BibitemOpen
\bibfield  {author} {\bibinfo {author} {\bibfnamefont {S.}~\bibnamefont
		{Proti{\`e}re}}, \bibinfo {author} {\bibfnamefont {A.}~\bibnamefont
		{Boudaoud}},\ and\ \bibinfo {author} {\bibfnamefont {Y.}~\bibnamefont
		{Couder}},\ }\bibfield  {title} {\bibinfo {title} {Particle-wave association
		on a fluid interface},\ }\href@noop {} {\bibfield  {journal} {\bibinfo
		{journal} {J. Fluid Mech.}\ }\textbf {\bibinfo {volume} {554}},\ \bibinfo
	{pages} {85} (\bibinfo {year} {2006})}\BibitemShut {NoStop}%
\bibitem [{\citenamefont {Eddi}\ \emph {et~al.}(2009)\citenamefont {Eddi},
	\citenamefont {Fort}, \citenamefont {Moisy},\ and\ \citenamefont
	{Couder}}]{Eddi2009}%
\BibitemOpen
\bibfield  {author} {\bibinfo {author} {\bibfnamefont {A.}~\bibnamefont
		{Eddi}}, \bibinfo {author} {\bibfnamefont {E.}~\bibnamefont {Fort}}, \bibinfo
	{author} {\bibfnamefont {F.}~\bibnamefont {Moisy}},\ and\ \bibinfo {author}
	{\bibfnamefont {Y.}~\bibnamefont {Couder}},\ }\bibfield  {title} {\bibinfo
	{title} {Unpredictable tunneling of a classical wave-particle association},\
}\href@noop {} {\bibfield  {journal} {\bibinfo  {journal} {Phys. Rev. Let.}\
	}\textbf {\bibinfo {volume} {102}},\ \bibinfo {pages} {240401} (\bibinfo
	{year} {2009})}\BibitemShut {NoStop}%
\bibitem [{\citenamefont {Lighthill}(1978)}]{Lighthill1978}%
\BibitemOpen
\bibfield  {author} {\bibinfo {author} {\bibfnamefont {J.}~\bibnamefont
		{Lighthill}},\ }\href@noop {} {\emph {\bibinfo {title} {Waves in fluids}}},\
\bibinfo {edition} {1st}\ ed.\ (\bibinfo  {publisher} {Cambridge University
	Press},\ \bibinfo {year} {1978})\BibitemShut {NoStop}%
\bibitem [{\citenamefont {Bender}\ and\ \citenamefont
	{Orszag}(1999)}]{Bender1999}%
\BibitemOpen
\bibfield  {author} {\bibinfo {author} {\bibfnamefont {C.~M.}\ \bibnamefont
		{Bender}}\ and\ \bibinfo {author} {\bibfnamefont {S.~A.}\ \bibnamefont
		{Orszag}},\ }\href@noop {} {\emph {\bibinfo {title} {Advanced mathematical
			methods for scientists and engineers}}}\ (\bibinfo  {publisher} {Springer},\
\bibinfo {year} {1999})\BibitemShut {NoStop}%
\bibitem [{\citenamefont {Mol{\'a}{\v c}ek}\ and\ \citenamefont
	{Bush}(013b)}]{Molacek2013}%
\BibitemOpen
\bibfield  {author} {\bibinfo {author} {\bibfnamefont {J.}~\bibnamefont
		{Mol{\'a}{\v c}ek}}\ and\ \bibinfo {author} {\bibfnamefont {J.}~\bibnamefont
		{Bush}},\ }\bibfield  {title} {\bibinfo {title} {Drops walking on a vibrating
		bath: towards a hydrodynamic pilot-wave theory},\ }\href@noop {} {\bibfield
	{journal} {\bibinfo  {journal} {J. Fluid Mech.}\ }\textbf {\bibinfo {volume}
		{727}},\ \bibinfo {pages} {612} (\bibinfo {year} {2013b})}\BibitemShut
{NoStop}%
\bibitem [{\citenamefont {Gonzalez}\ \emph {et~al.}(1989)\citenamefont
	{Gonzalez}, \citenamefont {McCluskey}, \citenamefont {Castellanos},\ and\
	\citenamefont {Barrero}}]{Gonzalez1989}%
\BibitemOpen
\bibfield  {author} {\bibinfo {author} {\bibfnamefont {H.}~\bibnamefont
		{Gonzalez}}, \bibinfo {author} {\bibfnamefont {F.}~\bibnamefont {McCluskey}},
	\bibinfo {author} {\bibfnamefont {A.}~\bibnamefont {Castellanos}},\ and\
	\bibinfo {author} {\bibfnamefont {A.}~\bibnamefont {Barrero}},\ }\bibfield
{title} {\bibinfo {title} {Stabilization of dielectric liquid bridges by
		electric-fields in the absence of gravity},\ }\href@noop {} {\bibfield
	{journal} {\bibinfo  {journal} {J. Fluid Mech.}\ }\textbf {\bibinfo {volume}
		{206}},\ \bibinfo {pages} {545} (\bibinfo {year} {1989})}\BibitemShut
{NoStop}%
\end{thebibliography}
\end{document}